# A New Era in Human Factors Engineering: A Survey of the Applications and Prospects of Large Multimodal Models

LI Fan[1,2], LEE Ching-Hung[3*], HAN Su[1], FENG Shanshan[4,5], JIANG Zhuoxuan[6], SUN Zhu[4,5]

([1]Department of Aeronautical and Aviation Engineering, The Hong Kong Polytechnic University, Hong Kong; [2]Research Centre for Data Sciences & Artificial Intelligence; [3]School of Public Policy and Administration, Xi 'an Jiaotong University, China; [4]Centre for Frontier AI Research, A*STAR, Singapore; [5]Institute of High Performance Computing, A*STAR, Singapore; [6]Shanghai Business School)

**Abstract:** In recent years, the potential applications of Large Multimodal Models (LMMs) in fields such as healthcare, social psychology, and industrial design have attracted wide research attention, providing new directions for human factors research. For instance, LMM-based smart systems have become novel research subjects of human factors studies, and LMM introduces new research paradigms and methodologies to this field. Therefore, this paper aims to explore the applications, challenges, and future prospects of LMM in the domain of human factors and ergonomics through an expert-LMM collaborated literature review. Specifically, a novel literature review method is proposed, research studies of LMM-based accident analysis, human modeling and intervention design are introduced. Subsequently, the paper discusses future trends of research paradigm and challenges of human factors and ergonomics studies in the era of LMMs. It is expected that this study can provide a valuable perspective and serve as a reference for integrating human factors with artificial intelligence.

## 1 Introduction

Human factors engineering studies the interaction between humans and other elements of the systems, aiming to design systems that are more in line with human capabilities and limitations [1]. Human factors engineering originated in the aviation safety field during World War II and gradually expanded into areas such as workstation design, automobile manufacturing, interaction design, and healthcare [2]. Recently, with the development of artificial intelligence (AI), the research scope of human factors engineering is further expanding [3]. Specifically, large multi-modal models (LMMs) have demonstrated significant potential and broad application prospects in various fields such as industrial design and customer service [4], [5]. It is expected that the application of LMMs will gradually change the research paradigm of human factors engineering in three aspects, including human-machine interaction model, research subjects, and research methods.

Firstly, traditional human-machine interaction often focuses on the operation of physical devices such as control panels, buttons, touch screens, keyboards, and mice [6]. Recently, the emergence of large models enables machines/robots to understand and generate natural language, enabling more natural interactions [7].

For example, Huang et al. [8] adopted large language model to process free-form natural language and generate codes for robotic manipulation. Wang et al. [9] adopted large language model to achieve multi-modal human-robot interaction by providing high-level linguistic guidance. These new interaction methods enable people to interact with machines more naturally and intuitively, greatly enhancing the convenience and efficiency of human-machine interaction.

Secondly, LMM-enabled intelligent agents, also named as AI Agents is becoming new research focuses on the field of human factors engineering. According to a survey published in 2024 [7], there is a surge in research on LMM-based autonomous agents. AI Agents differ significantly from traditional mechanical systems. Traditional human factors engineering typically focuses on the how humans interact with static and predictable systems, such as factory assembly lines or aviation flight control systems. These systems are relative stable, and their behaviors are predictable. In contrast, AI Agents are more dynamic and complex, they able to learn, adapt, and make autonomous decisions[7]. In addition, the behavior and decisions of intelligent agents are often influenced by multiple factors, including user instructions, contextual factors, and the evolution of the system itself [10]. Therefore, human-AI Agents interaction design needs to consider not only the needs and behavioral patterns of users but also the autonomy characters of AI Agents [10]. Hence, methods and principles of human factors engineering may need to be re-evaluated and adjusted to adapt to new human-AI Agents interaction paradigms and challenges.

Lastly, LMMs also offer new research methods for human factors engineering. For instance, traditional human factors engineering often recruited some participants even professional subjects in user experiments to collect user data [11], [12]. While LMMs increasingly have the capability to simulate human-like responses and behaviors, they can be used for large-scale and rapid testing of theories and hypotheses related to human behaviors [5]. In addition, traditional data analysis methods typically employ statistical analysis, machine learning, deep learning, etc **Error! Reference source not found.**. For example, in the field of vigilance and fatigue monitoring, a research study developed a novel fatigue detection model, demonstrating up to 89% accuracy in case studies at maritime traffic service centers [13]. Nevertheless, these methods suffering from explainability and reasoning issues [14]. In recent years, several studies have begun to adopt LMMs in analyzing human behavior and psychophysiological data [15], [16]. These techniques are capable of processing large volumes of high-dimensional, unstructured data, uncovering complex patterns and hidden relationships within the data, thereby delivering in-depth and reasonable analysis. However, how to ensure transparency and replicability is still a significant challenge.

In summary, the integration of LMMs and human factors engineering represents a new research field with both challenges and opportunities. Therefore, conducting in-depth exploration of the applications, challenges, and prospects of LMMs in the field of human factors engineering can provide significant theoretical and practical contributions. This paper aims to achieve this exploration through an expert-LMM collaborated literature review approach. Section 2 presents the literature review methodology. Section 3 reviews the existing human factor studies with the application of LMMs, including LMM-based accident analysis, human modelling, and intervention design. Section 4 discusses how LMMs reshape human factors research paradigm. In addition, the challenges and prospects of LMMs in the field of human factors engineering research are discussed in Section 4. The last section concludes this study and points out the limitations of this study.

## 2 Methods

Considering that research studies on LMMs in the field of human factors engineering are relatively limited, and most relevant literature is published on the arXiv preprint platform, we conducted literature collection and analysis via Google Scholar using a systematic review method. Traditionally, literature reviews are conducted manually by experts, which is time-consuming and inefficient. This study proposes an expert-LMM collaborative method for conducting the literature review. As shown in Figure 1, experts initially generated several keywords based on their knowledge, and then the LMM refined these keywords to generate more suitable ones. First, the expert conducted an analysis of the traditional research process in human factors engineering to identify potential applications of LMMs in this field. Traditionally, human factors engineering studies implement task analysis, statistical analysis, and accident analysis to evaluate system performance and identify potential problems, and then design solutions from five aspects, including training, human selection, equipment, task, and environment [17]. It is expected that LMMs can facilitate several parts of the classical process of human factors engineering and may even reshape this process. Hence, the initial codes are defined based on the classical process of human factors engineering. ChatGPT 4 Turbo was adopted to refine the codes. The coding process is iterative, including identifying common themes, comparing and contrasting findings, and redefining and updating the codes.

Through multiple iterations, the keywords used for searching mainly included "LMM/LLM/GPT" combined with "human factors/accident analysis/EEG/error/interaction design/performance evaluation." These keywords aimed to collect as much literature relevant to the research topic as possible. During the search process, particular attention was paid to the titles and abstracts of the literature to determine their relevance to our research topic. Based on the literature analysis, the existing relevant literature was categorized into LMM-based accident analysis, human modeling, and intervention design.

With the collected articles, the experts conducted preprocessing and then proposed a multi-step prompting method for the literature review, as shown in Figure 2. The abstracts of the related articles were fed to ChatGPT 4 Turbo, which was instructed to conduct two steps with reflection. The self-reflection can greatly reduce LLM hallucination and improve the performance of the literature review [18]. Hence, we referred to the interactive self-reflection method [18] to conduct the expert-LLM collaborative literature review. Specifically, ChatGPT 4 Turbo was prompted to identify the research field of the collected abstracts and then identify the classical theory or research process of this field. The experts joined in the process to evaluate the results. Then, ChatGPT 4 Turbo separated the abstracts into several groups and summarized these abstracts. The summarization was evaluated from the consistency perspective [18]. If the consistency score was too low, the clustering and summarization process were repeated until the result was acceptable. Experts conducted the final check and revision.

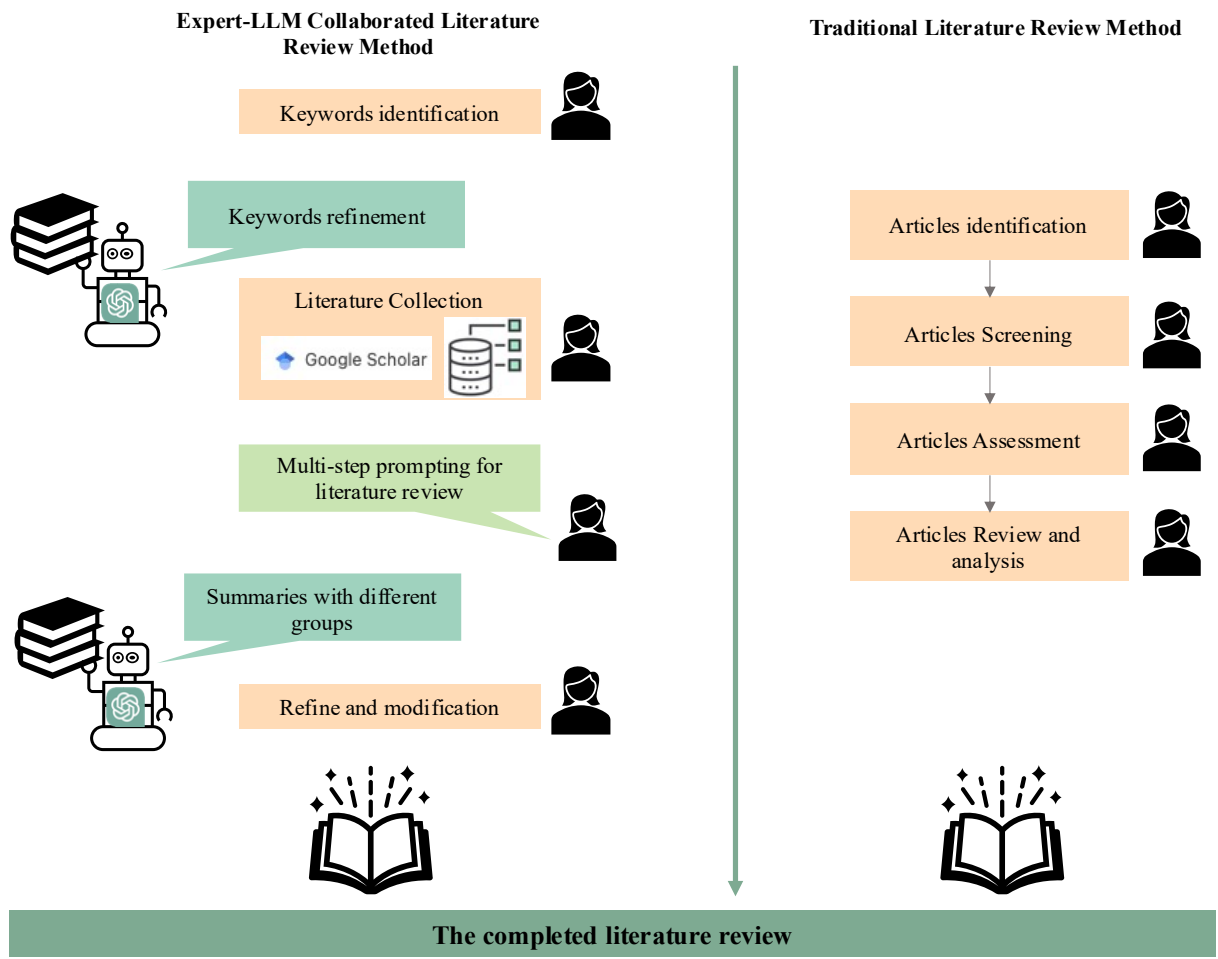

Figure 1: Comparing the proposed literature review method with traditional literature review method

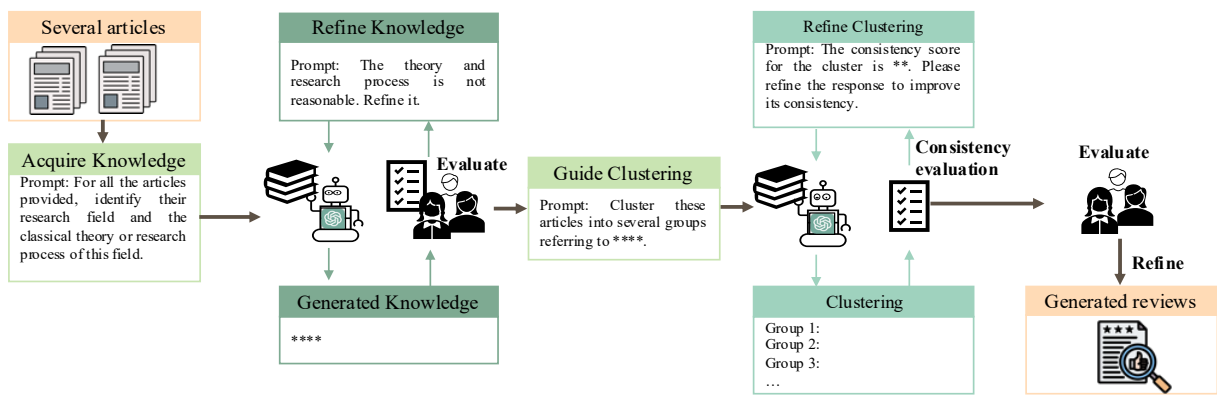

Figure 2: The multi-step prompting method for literature review.

## 3 Existing LMM-related Human Factors Studies

Traditional human-machine system performance analysis often includes task analysis, accident analysis, and statistical data analysis methods. In the era of LMMs, with the development of deep learning and natural language processing technologies, the capability for large-scale and diversified data analysis has been greatly enhanced. LMMs have demonstrated the ability to understand and generate complex human language, providing new possibilities for human factors engineering studies. This section reviews the existing studies on LMM-based accident analysis, human behavior modeling, and intervention design.

## 3.1 LMM-based Accident Analysis

Figure 3 shows the three clusters generated by ChatGPT 4 Turbo based on the articles provided by the authors. Each cluster represents a distinct aspect of the accident analysis process where LMMs are applied, from initial data extraction to in-depth multimodal analysis. We minorly refined the clusters and organized the literature into these three clusters, providing a structured overview of how LMMs contribute to different stages of the accident analysis process. The accident analysis normally includes data collection, data analysis, report generation, safety measurement implementation, and so on. The first clustered studies on LMM-based accident data collection mainly focused on narrative summarization and preprocess. These studies evaluate the effectiveness of LMMs in summarizing accident narratives and identifying human factors, which are crucial for understanding the underlying causes of accidents. For example, study [18] assess ChatGPT's ability to generate accurate summaries of accident events and identify human factors issues. Study [19] examines LMMs' performance in answering binary and complex questions derived from traffic accident narratives.

For data analysis, existing studies focus on the use of LMMs to extract key information from textual data and categorize it based on specific criteria. Extracting key information from a large number of accident reports and witness narratives is the primary step in accident analysis. To deal with lengthy and non-standardized accident data, some studies have tried to use LMMs to extract key information [20], [21]. For example, Ehsan Ahmadi et al. compared several large language models (such as GPT-3.5, GPT-4, and Gemini Pro) in extracting key information from construction accident reports [20]. The key information includes accident causes, injury reasons, affected body parts, accident severity, and accident time. They found that GPT-4 achieved higher accuracy on most aspects, especially in "accident severity" and "accident time". In the classification of "injury reasons," Gemini Pro demonstrated superior performance. Andrade and Walsh fine-tuned the BERT model using aviation safety reports, resulting in the SafeAeroBERT model [22]. They adopted this model to classify reports into four distinct classes based on accident causal factors, exhibiting superior performance in classifying reports associated with weather and procedural factors. Kierszbaum et al. introduced the ASRS-CMFS model to conduct diverse classifications based on abnormal types of causal events [23].

The final cluster of the existing studies covers research on the advanced application of LMMs in processing multimodal inputs and their integration into broader accident analysis workflows. Grigorev et al. explored the integration of large models into machine learning workflows for accident management, simplifying the process of feature extraction from unstructured text data[24]. Furthermore, studies demonstrating the use of LMMs to analyze and integrate various data types (images, text, videos, audio) for a more comprehensive accident analysis, such as AccidentGPT's unified analysis framework [25] and multisensor perception frameworks for traffic accident analysis [26].

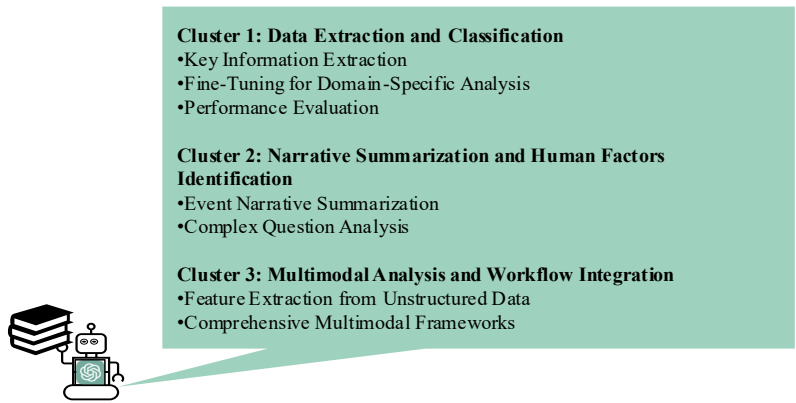

Figure 3: Three clusters of studies on LMM-based accident analysis (Generated by ChatGPT 4 Turbo)

## 3.2 LMM-based Human Modelling

LMMs provided a new direction for conducting user experiment and analyzing human behaviors [7]. Nevertheless, there are limited studies on LMM-based human modelling. Hence, we did not adopt the proposed multi-step prompting method in this section. The expert-based summarization is presented below.

Several studies explored the possibility of modelling human behaviors with LMMs. For example, Zhang et al. applied LMM as a zero-shot human model to predict behavior in human-computer interaction [27]. They tested the zero-shot human model across social datasets and confirmed that it can simulate complex human behavior. Additionally, Abbasiantaeb et al. [28] further confirmed LMMs' advantages in understanding and generating complex behavioral patterns through simulating interpersonal dialogues. A study applied LMMs to simulate social behavior in specific tasks, revealed its adaptability in handling complex social interactions and task coordination [29]. These studies validate the possibility of using LMM to simulate human behavior for further applications in user experimentation and personnel behavior analysis.

Utilizing physiological data, such as eye-tracking and electroencephalogram (EEG) data, to predict and monitor human physiological states, has become a widely researched topic. The existing studies normally adopt machine learning or deep learning techniques to predict human fatigue, workload, stress, vigilance, and situational awareness [13], [30], [31], [32], [33]. Nevertheless, these methods suffer from explainability issues. It is hard to understand the causal relations and the reasoning process of machine/deep learning algorithms. The rapid advancement of LMMs provides a potential new direction for explainable eye movements and EEG data analysis. Therefore, some studies tried to utilize the latest advances in LMMs and proposed EEG-GPT[15], a comprehensive EEG classification method. EEG-GPT achieves excellent performance in the few-shot learning paradigm, comparable to the current state-of-the-art deep learning methods, using only 2% of the training data to distinguish between normal EEG and abnormal EEG. Additionally, it possesses unique advantages in providing analysis of intermediate reasoning steps.

Figure 4 shows the possible topics of human modelling/simulation. The existing studies are mainly focus on modelling or simulate human social behaviors, while limited studies can be found in physical and cognitive spaces.

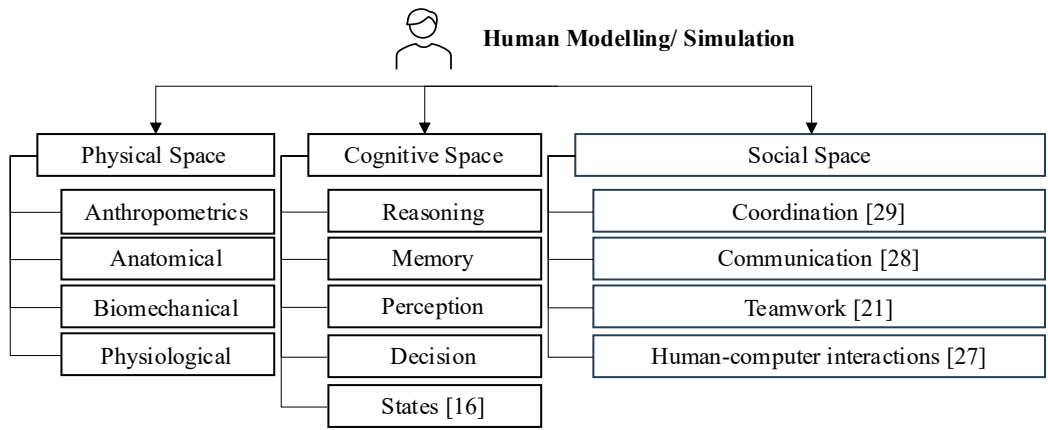

Figure 4：Human Modelling/ Simulation with LMMs

## 3.3 Interventions Design

Human error preventions strategies normally include task design, environment design, interface design, human selection and training design. The multi-step prompting method was adopted to conduct the initial clustering. Two clusters are generated as shown in Figure 5. Some studies have tried to apply LMMs to improve the human error preventions strategies or even design human error preventions from a new perspective. The existing studies can be classified into two types, LMM-based new design tools for experts and for users [34], [35], as shown in Figure 6.

For experts, several studies tried to apply LMM to facilitate new functions and assist experts. For example, in 2022, Google found that using LMM for prompt-based prototyping greatly enhanced communication among project managers, designers, and developers, speeding up the prototyping process[36], [37]. In 2024, Zheng et al. [38] fine-tuned a general large language model using annotated datasets and error-assisted iterations, enabling robots to effectively understand query statements and generate specific programming code for the manufacturing industry.

For users, the application of LMMs can make human-computer interaction experiences more natural and facilitate seamless interactions. Leveraging their ability for contextual understanding and coherent text generation, LMM-enabled products and services can provide instant and more human-like feedback, which aligns better with natural human habits[39]. A study applied LMM to achieve empathic mental inference, making user interface (UI) and user experience (UX) design more human-centered and better meet users' personalized needs[35]. In 2023, a psychological counseling assistant tool was developed based on LMMs, which provided timely feedback to psychological patients' inquiries, effectively reducing their stress [40]. LMMs are also integrated with robots, enhancing both verbal and non-verbal interactions between users and robots, thus making engagement more comprehensive and natural Error! Reference source not found.. Meanwhile, through multi-agent collaboration mechanisms, LMMs are expected to better simulate interactions in complex human-machine system environments, thereby aiding in understanding the division of tasks among different roles, reducing human errors in complex environments [41].

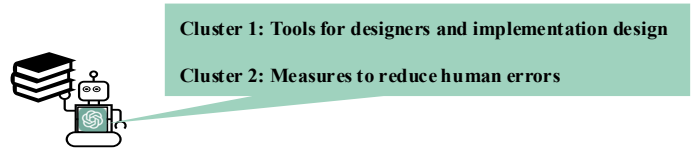

Figure 5: Two clusters of studies on LMM-based intervention design (Generated by ChatGPT 4 Turbo)

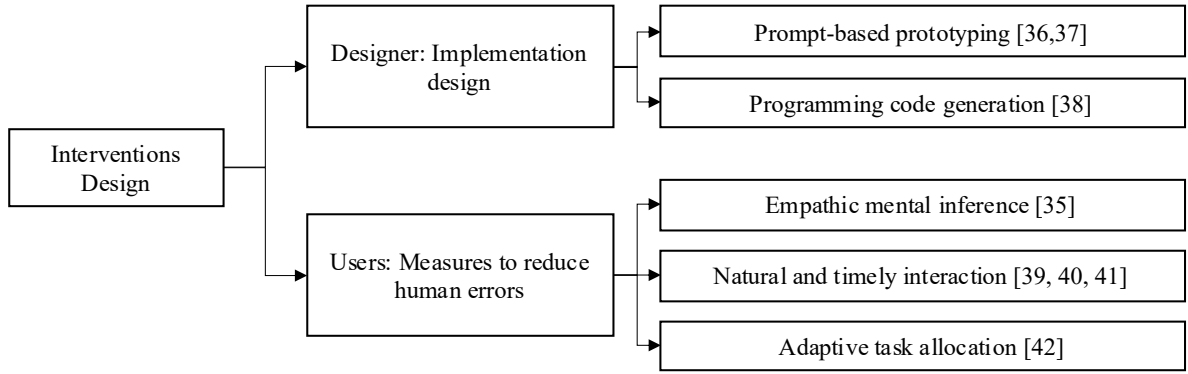

Figure 6: Interventions design for reducing human errors

## 4 LMMs-reshaped Human Factor Research Paradigm

This section discusses how LMMs reshape human factor studies from three aspects, namely research objects, system performance evaluation, and human error interventions design.

### 4.1 Research Objects: From Automation to Autonomy to Intelligence

The main research objects of human factor studies are interactions between human and other elements of the specific system[17]. At the beginning, human factors studies focused on fixed and static mechanical and electronical systems, such as workspace, chairs, and hand tools. In the recent years, we paid more attention to automation, such as autopilots, autonomous driving, unmanned aerial vehicles, and autonomous robots [42], [43]. In the near future, the advanced artificial intelligent techniques are driving towards AI agents [10]. Compared to traditional machinery, AI agents possess unique characteristics such as autonomy and the capacity for evolution. They are capable of reflection, planning, tool usage, and taking action [7], [10]. As shown in Figure 7, historically, the focus has been on human-machine interactions. This evolved into an emphasis on human-computer interactions. Moving forward, the spotlight will increasingly shift towards human-AI interactions, underscoring the advanced capabilities and complexities introduced by artificial intelligence.

Human-AI interaction introduces unique challenges and opportunities compared to traditional human-machine interactions and human-computer interaction. While there is overlap with some research topics, such as interface design and interaction design, there are also distinct areas of focus specific to human-AI interactions. First, considering the autonomy of AI agent, how to achieve effective human-AI shared decision-making is still waiting to be studied. This involves studies on human expertise assessment, user input analysis, and mutual understand between human and AI agents. Second, similar to existing investigation on the effects of automation level, we need to do more research to investigate the effects of different levels and gradations of autonomy on human-AI interactions. Third, besides the fixed levels of autonomy, the effects of adaptive and dynamic autonomy should be studied, too. Forth, the ethical and social impacts of AI agents will be an important research topic.

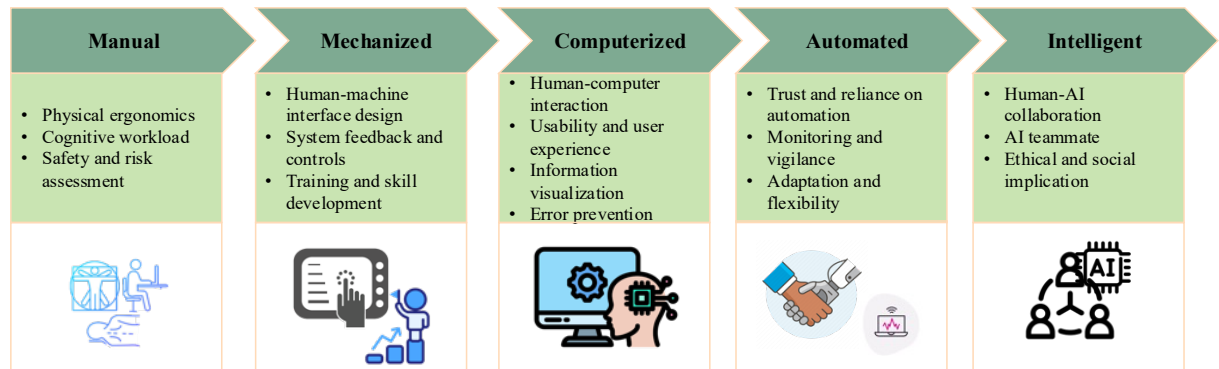

Figure 7: Human factors research subjects across different phases of technological development

## 4.2 System Performance Evaluation: From Accessible to Acceptable to Sustainable

The roots of human factors studies can indeed be traced back to physical ergonomics, initially focused on optimizing the design of tools, equipment, and workspaces to fit the physical capabilities and limitations of human users. Physical ergonomics, in particular, emphasized the importance of factors such as posture, movement, and workspace layout in promoting comfort, safety, and efficiency for workers performing manual tasks. One of the most important system evaluation metrics of physical ergonomic is accessibility[44].

As the field of human factors evolved, its scope was expanded to human-system interaction, including more aspects of human cognitive. Cognitive ergonomics focuses on understanding human information processes, such as perception, attention, memory, decision-making, and action [45].

Recently, with the raise of AI agent, a new perspective of system performance evaluation considering more factors besides physical and cognitive ergonomics is necessary. Firstly, the behavior of LMMs may have a certain level of uncertainty, which could affect the stability and predictability of system performance. The impact of the instability and unpredictability of LMMs on human and system performance remains unclear. For example, human-to-human interaction often exhibits high fault tolerance, but machine errors can easily lead to a collapse in human trust to the machine. Investigating how errors in LMMs affect human performance is worthwhile. Secondly, LMMs applications differ significantly from traditional mechanical systems; they are prone to errors but can be quickly corrected through real-time interaction. Therefore, the interaction between humans and LMMs applications face new problems. Both human and AI systems exhibit dynamic self-correction and ambiguity, necessitating further research into system performance evaluation. Finally, the application of LMMs may have long-term impacts, such as changes in user behavior and risks to data privacy, which may be challenging to observe and evaluate in the short term.

Considering that LMM-based systems are changing the role of the specific system from assistive tool to teammate, we propose to evaluate the human-AI system performance by referring to human relationship evaluation theories. According to the Maslow's hierarchy human needs, we propose sustainable human-AI interactions. The sustainable human-AI interactions refer to the design, implementation, and use of artificial intelligence (AI) systems in a manner that promotes long-term benefits for both humans and the AI systems

[4]. It encompasses several dimensions, such as environmental sustainability, social sustainability, economic sustainability, culture and ethical sustainability. In this study, we mainly focus on human factors-related aspects, namely social sustainability. Social sustainability means that human and AI can keep a long-term social relationship. As illustrated in Figure 8, the evolution of human factors research aligns with Maslow's hierarchy of human needs. Initially, the field concentrated on physiological requirements, leading to a focus on the physical ergonomics and accessibility of systems. As the foundational needs are met, the emphasis shifts to ensuring safety, expanding the scope of human factors to include cognitive aspects such as the acceptance and usability of technology. Hence, advancing further up Maslow's hierarchy, we expect that human-AI interactions should address social, esteem, and self-actualization needs in the near future. Recognizing and fulfilling these higher-level needs is crucial for cultivating a sustainable and mutually beneficial relationship between humans and AI systems.

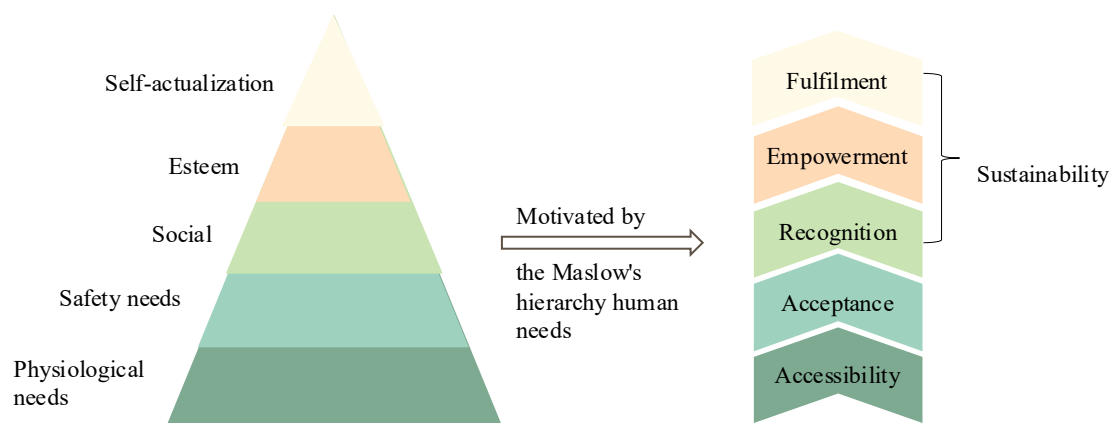

Figure 8: The Maslow's hierarchy human needs-driven sustainable human-AI interactions

### 4.3 Human Error Management: From Offline to Online and Inline Management

The interaction between humans and large model applications will pose a new challenge, as LMMs normally exhibit dynamic instability and ambiguity. The dynamic interactions between human and LMMs may trigger new patterns of human errors. As shown in Figure 9, human errors include knowledge-based mistakes, rule-based mistakes, skill-based errors[46]. Both knowledge-based mistakes and rule-based mistakes resulted from misinterpretation of the situations. With the applications of LMMs, more and more generated content will appear in real-time interactions, leading to unintended misinterpretations or the dissemination of misleading information. The generated content may be wrong but difficult to distinguish, leading to increasing in knowledge-based mistakes and rule-based mistakes.

Skill-based errors include slips and lapse. The applications of LMMs can be adaptive, whereas human beings may not be accustomed to adaptive interactions, potentially increasing the risk of skill-based errors, namely habit-based slips. Dynamic interface can reduce the risks of establishing habits and then reduce the habit-based slips. Nevertheless, users are not familiar with the interface may results in slips and lapse, too. So, how to balance the dynamic and familiarity to reduce the risks of human errors is an important human factor research topic.

Though LMMs may bring new challenges in inducing human errors, they also provide the opportunities in managing human errors. As shown in Figure 9, we can provide corresponding strategies to avoid or correct human errors. Specifically, to reduce knowledge-based mistakes, LMMs applications can generate more related and useful information to enhance users' situation assessment. To avoid rule-based mistakes, we can develop customized assistance based on users experience and habit to provide adaptive information. In addition, user intentions of actions can be monitored and corrected if necessary. For the errors of slips and lapses, LMMs can generate suggestions or warnings in real-time, by monitoring users' intentions and behaviors. Overall, comparing with offline or post-accident management, such as accident analysis and user complains, more and more online monitoring and in-line human error management will be achieved in the near future.

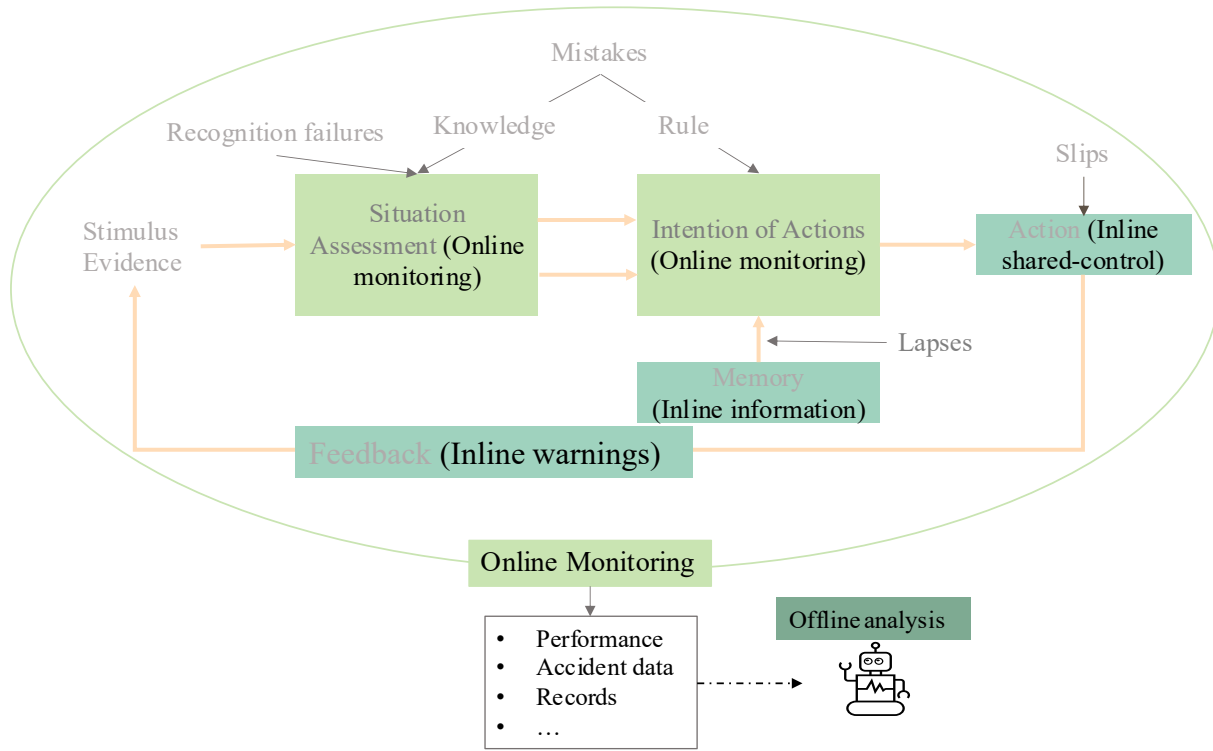

Figure 9: LMMs reshape human error management

## 5 Conclusion

In summary, the widespread application of LMMs in various fields such as healthcare, social psychology, and industrial design in recent years has provided new directions and opportunities for human factors and ergonomics research. From automated machines to systems with autonomy and evolution capability, LMMs have expanded the scope of human factors research and provided new avenues for development. Due to the limited existing literature, this paper adopts a combined approach of systematic literature review and commentary. Based on classical process of human factors research methods, it explores the application, challenges, and future trends of LMMs in the fields of human factors.

Existing literature mainly focuses on accident analysis, human behaviour modelling, and intervention design. LMMs' capabilities in language analysis and reasoning enhance accident factors identification and accidents severity prediction. In addition, LMMs assist experts and designers in rapidly generating design concepts and optimizing human-machine interaction to enhance user experience.

The existing LMMs literatures in human factors are limited. Hence, we integrated research experience and existing studies to analyse the development direction of human-machine ergonomics in the era of LMMs from three aspects: research objects of human-machine ergonomics, performance evaluation of human-machine systems, and design of strategies for human error prevention. Overall, the field of human-machine ergonomics will face and address increasingly dynamic instability and ambiguity.

## Acknowledgement:

The study received partial support from the Hong Kong Polytechnic University (P0038933 and P0038827) and was also partially funded by the Research Centre for Data Sciences & Artificial Intelligence (P0046166). The draft was proofread by ChatGPT 4.0 Turbo.